%% file: paper.tex
\documentclass{article} 
\usepackage{iclr2021_conference,times}

\input{math_commands.tex}

\usepackage{hyperref}
\usepackage{url}

\usepackage{graphicx}
\usepackage{subfig}

\title{Detection of COVID-19 Disease using Deep Neural Networks with Ultrasound Imaging}

\author{Carlos Rojas-Azabache, Karen Vilca-Janampa \hspace{2cm} \\
Universidad Nacional Mayor de San Marcos, Peru \\
\texttt{\{carlos.rojas6,10140251\}@unmsm.pe} \\
\And 
Renzo Guerrero-Huayta, Dennis Núñez-Fernández \hspace{2cm} \\
Universidad Nacional de Ingeniería, Peru \\
\texttt{\{rguerreroh,dnunezf\}@uni.pe} \\
}

\iclrfinalcopy 
\begin{document}

\maketitle

\begin{abstract}
The new coronavirus 2019 (COVID-2019) has rapidly become a pandemic and has had a devastating effect on both everyday life, public health and the global economy. It is critical to detect positive cases as early as possible to prevent the further spread of this epidemic and to treat affected patients quickly. The need for auxiliary diagnostic tools has increased as accurate automated tool kits are not available. This paper presents a work in progress that proposes the analysis of images of lung ultrasound scans using a convolutional neural network. The trained model will be used on a Raspberry Pi to predict on new images.
\end{abstract}

\section{Introduction}

COVID-19 is the most recently discovered infectious disease caused by the coronavirus. Both this new virus and the disease it causes were unknown before the outbreak in Wuhan, China, in December 2019. Today, COVID-19 is a pandemic affecting many countries around the world. It has had a devastating effect on daily life, public health and the global economy. The current COVID-19 pandemic has impacted the world with over 18.35 million infections and over 6,96,147 deaths so far (as of August 5, 2020) \citep{1}.  Early identification, isolation and care of patients is a key strategy for optimal management of this pandemic \citep{2} and therefore a new approach can have a positive impact on the population and on the different remote areas of the country.

Since the appearance of cases of COVID-19, several methodologies based on machine learning have been developed for its detection using medical images, gaining more relevance in the pulmonary affectation those of the chest X-ray, computed tomography (CT) scanner and ultrasonography type. Such approaches have shown favorable results. In \citep{3} transfer learning is used with X-ray, ultrasound and CT images, and using the model VGG19 with an accuracy of up to 86\% for chest radiography, 100\% for ultrasound and 84\% for CT. In \citep{4} they make use of 1103 ultrasound images of healthy patients with COVID-19 and bacterial pneumonia and a CNN POCOVID-Net, giving a sensitivity of 0.96, a specificity of 0.79. At \citep{5} they use ultrasound imaging and a novel deep network, derived from Spatial Transformer Networks, that predicts the quantification of disease severity and provides the location of pathologic features in a weakly monitored manner.

However, the benefits and risks of each modality of imaging methods (chest radiography, CT, and ultrasonography) will depend on the patient and the stage of disease progression. In the current state, CT is the imaging method of choice for COVID-19 pneumonia imaging, which is characterized by ground-glass opacity (GGO) abnormalities at the onset of the disease, followed by the crazy-paving pattern, and finally consolidation in the more advanced stage of the disease \citep{6} \citep{7}. Although chest CT is useful, it is expensive and not available in many institutions, and sterilization of the CT scanner after use can cause delays in the care of other patients \citep{8}. Another method used is the chest X-ray, but when using radiation it requires certain requirements regarding permits and the construction of the environment that it needs to function, it is necessary to comply with regulations, as well as personnel with a license to use radiation and also good maintenance of the equipment, without considering that in the country this procedure is used only for control and not for detection.  

In this paper we propose a convolutional neural network in grayscale images obtained from lung ultrasound to classify whether the image belongs to a patient with COVID-19 or a healthy one. The system will perform inference on a Raspberry Pi embedded computer to work in different health centers with or without an Internet connection.

\section{Methodology}

\begin{figure}[h]
  \centering{\includegraphics[width=120mm]{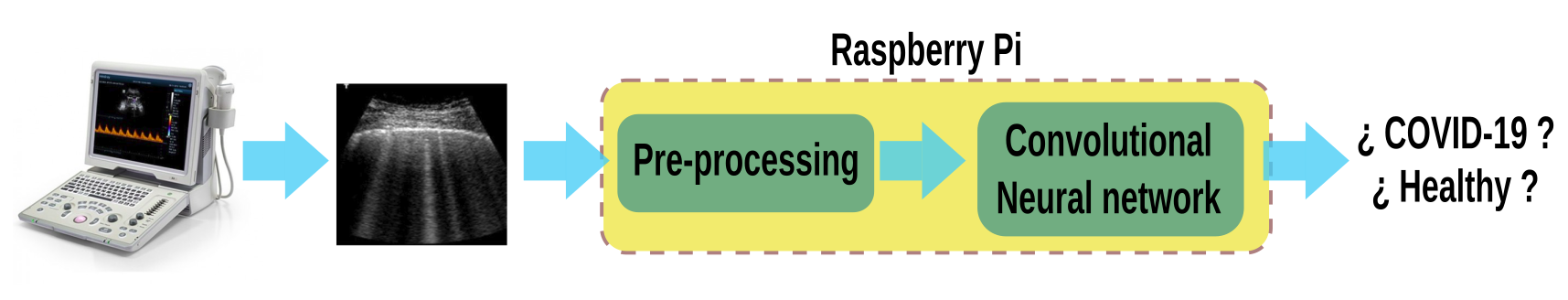}}
  \caption{Diagram for the proposed system}
  \label{image1}
\end{figure}

In this work we will use lung ultrasound images to perform the detection of COVID-19, as it is a low cost and high availability method that does not use ionizing radiation. In addition, we propose the use of a convolutional neural network to perform the classification task. The trained model will be used on a Raspberry Pi to predict on new images with fast response time and without Internet access, see Fig.~\ref{image1}.

The dataset used for this work was collected from different private clinics in the town of Lima, and extracted with various brands of portable ultrasound machines. The proposed number of videos for collecting is 50, from which a total of 750 images will be extracted for cases with COVID-19 and 750 images for healthy cases. The images will be presented in grayscale format and with a size of 512x512 pixels.

The proposed architecture is based on the stages of image processing, data enhancement, model construction and finally a classification. In the processing stage, a scaling to 224x224 pixels will be performed, and also data augmentation will be used, thus obtaining 10 times the initial amount \citep{9}. In the construction of the model, it has been taken into account previously proposed architectures for the classification of images from chest X-rays such as VGG-16 \citep{10} and the COVIDX-Net based on seven different architectures from DCNN, VGG19, DenseNet201, InceptionV2, Resnet101, InceptionV3, Xception and MobileNetV2 \citep{11}. In addition, for the construction of the model, transfer learning from the POCOVID-Net network will be used \citep{4}, training the dense layers so that they can classify the lung ultrasound images that are in the dataset. Therefore, the proposed architecture will have the following composition: 2xC(150x150x64) - MP(75x75x64) - 2xC(75x75x128) - MP(37x37x128) - 3xC(37x37x256) - MP(18x18x256) - 3xC(18x18x512) - MP(9x9x512) - 3xC(9x9x512) - MP(4x4x512) - F(8192) - FC(2). Where, C: Conv. layer, F: Flatten, FC: Full connection, MP: Max Polling.

\section{Conclusions}

In this work we propose the use of lung ultrasound imaging and a convolutional neural network for the detection of COVID-19, and the use of the trained model in a Raspberry Pi to perform prediction on new images. Ultrasonography, since it does not use ionizing radiation, decreases the cost and has high availability. It can also be performed several times to evaluate the patient's condition, an alternative that is not possible with chest X-rays or tomography. Despite our work is interesting, we are still collecting the images. In this way, the project will have a positive impact on the population and on the different remote areas of the country, since it will be of great support for the diagnosis, prognosis, monitoring, recovery and complications that Covid-19 patients present.

\clearpage

\bibliography{references}
\bibliographystyle{iclr2021_conference}

\end{document}

%% file: math_commands.tex
\usepackage{amsmath,amsfonts,bm}

\def\eqref#1{equation~\ref{#1}}

\def\1{\bm{1}}

\DeclareMathAlphabet{\mathsfit}{\encodingdefault}{\sfdefault}{m}{sl}
\SetMathAlphabet{\mathsfit}{bold}{\encodingdefault}{\sfdefault}{bx}{n}

